\documentclass[12pt]{article}
\usepackage{amssymb,amsmath,graphicx}
\usepackage{float}

\usepackage{epsfig}

\usepackage{times}
\usepackage[T1]{fontenc}

\usepackage{array}
\usepackage{picture}
\usepackage{subfig}
\usepackage{color}

\usepackage[margin=15pt]{caption}

\def\bc{\begin{center}}
\def\ec{\end{center}}
\def\beq{\begin{equation}}
\def\eeq{\end{equation}}
\def\bea{\begin{eqnarray}}
\def\eea{\end{eqnarray}}

\def\bdm{\begin{displaymath}}
\def\edm{\end{displaymath}}
\def\ds{\displaystyle}

\newcommand{\gsim}{~{}_{\textstyle\sim}^{\textstyle >}~}
\newcommand{\lsim}{~{}_{\textstyle\sim}^{\textstyle <}~}

\font\ku=cmti12 scaled\magstep2

\font\ku=cmti9 

\begin{document}

\author{{V.~G.~Bornyakov$^{a,b,c}$ and R.~N.~Rogalyov$^a$}\\[1mm]
{\ku a)   Institute for High Energy Physics NRC "Kurchatov Institute", 142281 Protvino, Russia}\\[-2mm]
{\ku b)  School of Biomedicine, Far East Federal University, 690950 Vladivostok, Russia}\\[-2mm]
{\ku c) Institute of Theoretical and Experimental Physics NRC "Kurchatov Institute",}\\[-2mm]
{\ku 117259 Moscow, Russia}\\
}

\title{Gluons in two-color QCD at high baryon density}


\maketitle

\abstract{Landau gauge  longitudinal and  transverse gluon propagators are studied in lattice QCD with gauge group $SU(2)$ at varying temperature and quark density. In particular, it is found that the longitudinal propagator decreases with increasing quark chemical potential at all temperatures under study, whereas the transverse  propagator increases with increasing  quark chemical potential at $T<200$~MeV and does not depend on it at higher temperatures. The relative strength of chromoelectric and chromomagnetic interactions is also discussed.
}


\section{Introduction}
We study Landau gauge gluon propagators in $N_f=2$ lattice QCD with gauge group $SU(2)$ at nonzero temperature and quark density. 
The details of the lattice action used to generate the gauge field configurations are described in  recent papers\cite{Bornyakov:2017txe, Astrakhantsev:2018uzd}. 

It is well known that lattice QCD approach is very successful at zero baryon density
but is inapplicable so far at high baryon density due to the so called sign problem \cite{Muroya:2003qs}.
It is then useful to study in the lattice regularization the theories similar to QCD (QCD-like) but without 
sign problem. In particular, QCD with $SU(2)$ gauge group \cite{Kogut:2000ek} (to be called below QC$_2$D) is one of such theories.  

QC$_2$D  was studied using various approaches which are also applicable to QCD at high baryon density. It is thus possible to check their predictions in the case of QC$_2$D confronting respective results with first principles lattice results.

The phase diagram of QC$_2$D in the $T-\mu_q$ plane is still not fixed solidly. 
In the studies of the $N_f=2$ lattice QC$_2$D with staggered fermionic action at
high quark density and $T=0$ it was demonstrated\cite{Bornyakov:2017txe} that the string tension $\sigma$ decreases
with increasing $\mu_q$ and becomes compatible with zero for $\mu_q$ above 850~MeV. In a more recent paper \cite{Boz:2019enj}, where $N_f=2$ lattice QC$_2$D with Wilson fermionic action
was studied, the authors did not find the confinement-deconfinement transition at low temperature. The reason for this discrepancy might be the lattice artifacts at large lattice values of $\mu_q a$.

In this paper we make a step toward the study of QC$_2$D phase diagram  in the $T-\mu_q$ plane. We concentrate on the study of the  Landau gauge gluon propagators at nonzero temperature and varying quark chemical potential and compare the results with our earlier findings obtained at zero temperature. 
We try to find the signs of the confinement-deconfinement transition in the 
gluon propagators dependence on the temperature and the quark chemical potential.

The gluon propagators are among important quantities to study, e.g. they play crucial role in the Dyson-Schwinger equations approach.
The $\mu_q$ and/or temperature dependence of the gluon propagators in Landau gauge  in lattice QC$_2$D were studied before in Refs.~\cite{Hands:2006ve,Boz:2013rca,Hajizadeh:2017ewa,Boz:2018crd}. 

One of the conclusions of Ref.~\cite{Boz:2018crd} was that, beyond
the hadronic phase in the $\mu_q-T$ phase diagram of QC$_2$D, the chromoelectric interactions are suppressed at low momenta, whereas the chromomagnetic - practically do not change.

The gluon propagators in QC$_2$D at nonzero $\mu_q$
were also studied in Ref.~\cite{Contant:2019lwf} 
with  help
of the Dyson-Schwinger equations approach and in  Ref.~\cite{Suenaga:2019jjv} using the
massive Yang-Mills theory approach at one-loop. The authors emphasize that after the agreement with the lattice results
for the gluon propagators will be reached their methods could be applied to real QCD at nonzero baryon density. Thus to provide unbiased lattice results is very important.

Our computations are completed on lattices 
$32^4, 32^3 \cdot 24, 32^3 \cdot 16, 32^3 \cdot 8$ 
with respective temperature values\footnote{Strictly speaking, for our symmetric $32^4$ lattice, $\frac{1}{N_t a}=140$~MeV. We consider this value as a good approximation for zero temperature in accordance with common practice.} $T=0, 188, 280, 560$~MeV 
and quark chemical potential $0 \leq  \mu_q < 1.8$~GeV. 
The lattice spacing for parameters used in this study was determined in
Ref.\cite{Bornyakov:2017txe} as $a=0.044$~fm. 
In this work we do not determine the line of
confinement-deconfinement transition in $T-\mu_q$ plane 
leaving this important task to future studies. It is pretty clear
that $T=560$~MeV is in the deconfinement phase for any value of $\mu_q$. Same is true for $T=280$~MeV, most probably. We expect that $T=188$~MeV is in the confinement phase at  $\mu_q=0$ and transition to the  deconfinement phase happens at high $\mu_q$.

\section{Gluon propagators}

The analysis of our data indicates that both the longitudinal and the transverse propagators are decreasing functions of the momentum; our data can be described using the fit function 
\beq\label{eq:GS_fit_widely_used}
D_{L,T}(p)= Z_{L,T}\;{1 + \delta_{L,T} \, p^2 \over p^4+2 R_{L,T} \, p^2 + M_{L,T}^2 }\; ,
\eeq
which, in particular, stems from the Refined Gribov-Zwanziger approach \cite{Dudal:2008sp}. It has received considerable attention in the literature
\cite{Dudal:2010tf,Cucchieri:2011ig,Oliveira:2012eh,Dudal:2018cli,Aouane:2011fv}.
The respective fitting procedure was considered in detail in Refs.~\cite{Bornyakov:2019jfz,Bornyakov:2020kyz} at $T=0$, here we describe the results for higher temperatures.
We use the usual normalization condition for the propagators, 
\beq\label{eq:norm_condition}
D^{ren}_{L,T}(\kappa^2)={1\over \kappa^2}\; ,
\eeq
at $\kappa=6$~GeV. Below we consider renormalized quantities omitting superscript 'ren'.
We consider only the soft modes $p_4=0$.

\subsection{Dressing functions}

\begin{figure}[H]

\vspace*{-30mm}
\hspace*{-18mm}\includegraphics[scale=0.4]{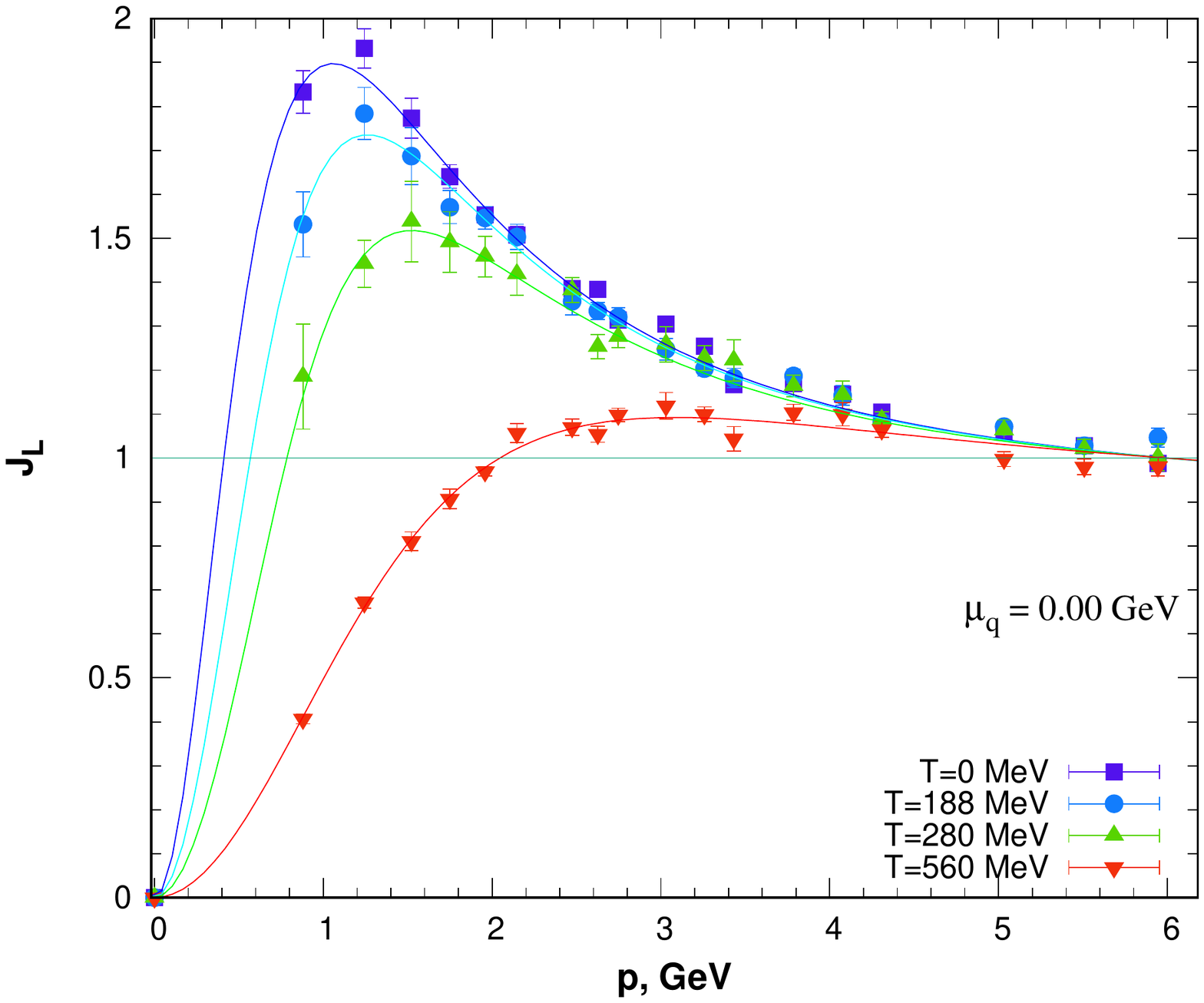}
\includegraphics[scale=0.4]{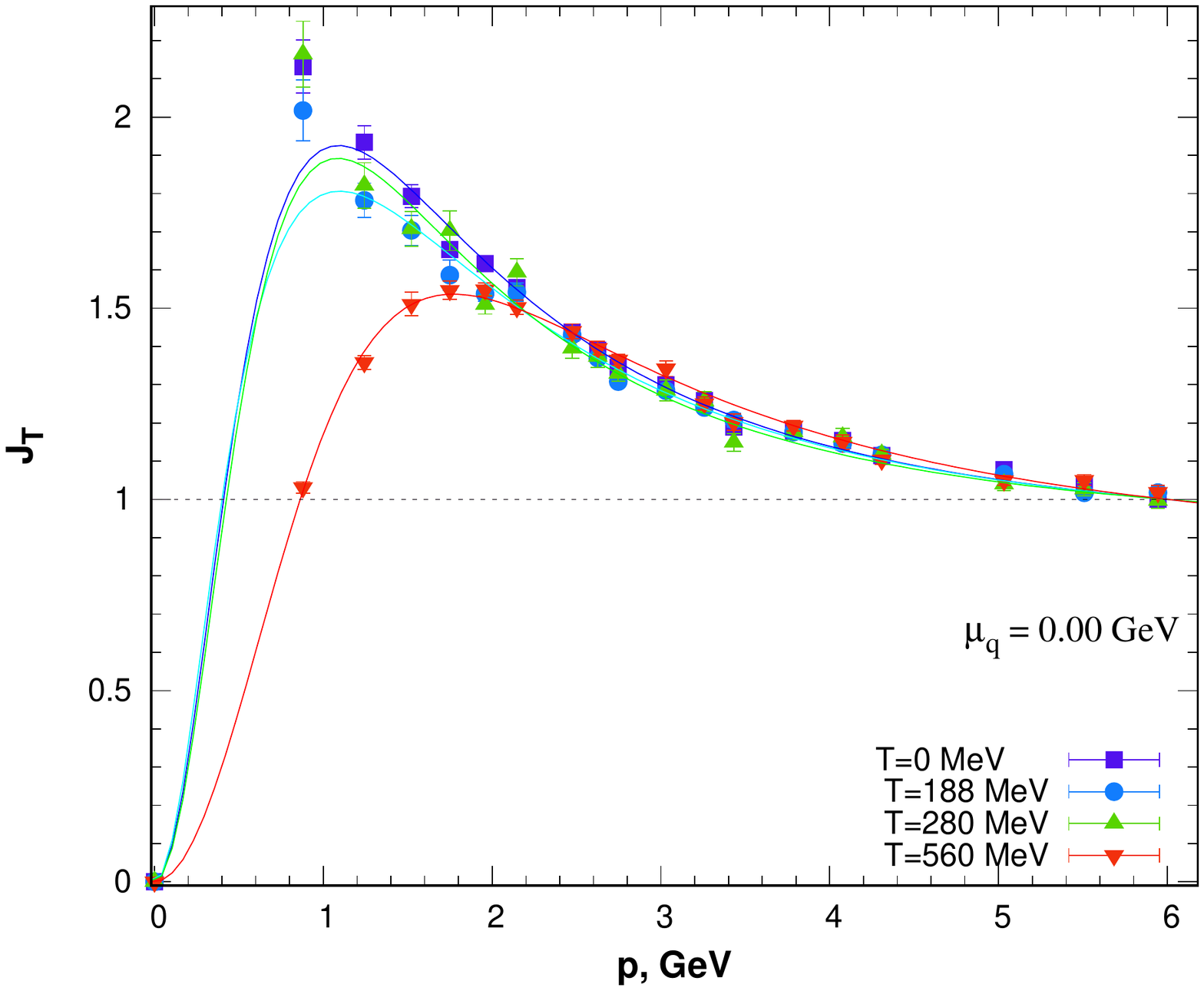}

\vspace*{-55mm}
\hspace*{-18mm}\includegraphics[scale=0.4]{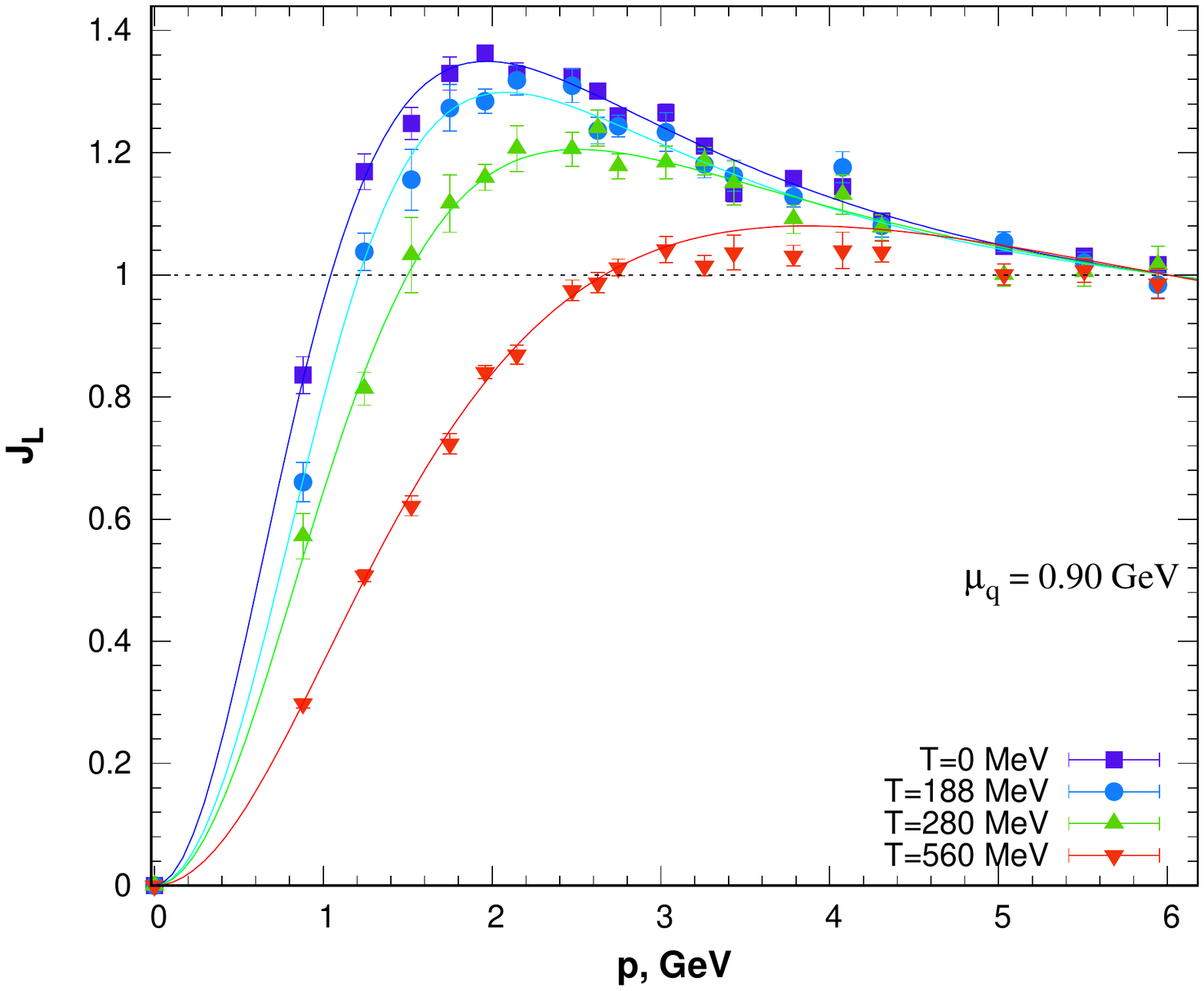}
\includegraphics[scale=0.4]{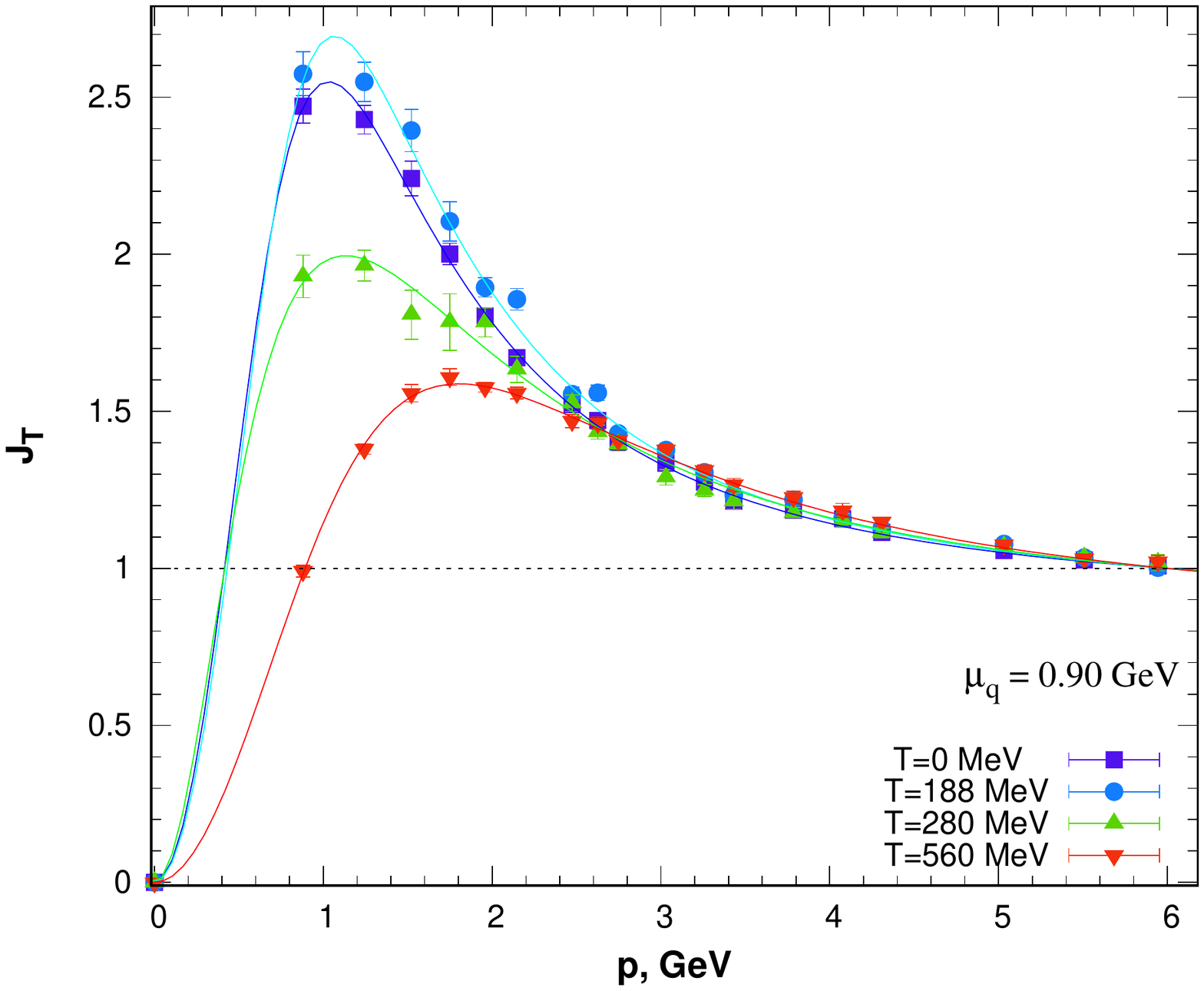}

\vspace*{-55mm}
\hspace*{-18mm}\includegraphics[scale=0.4]{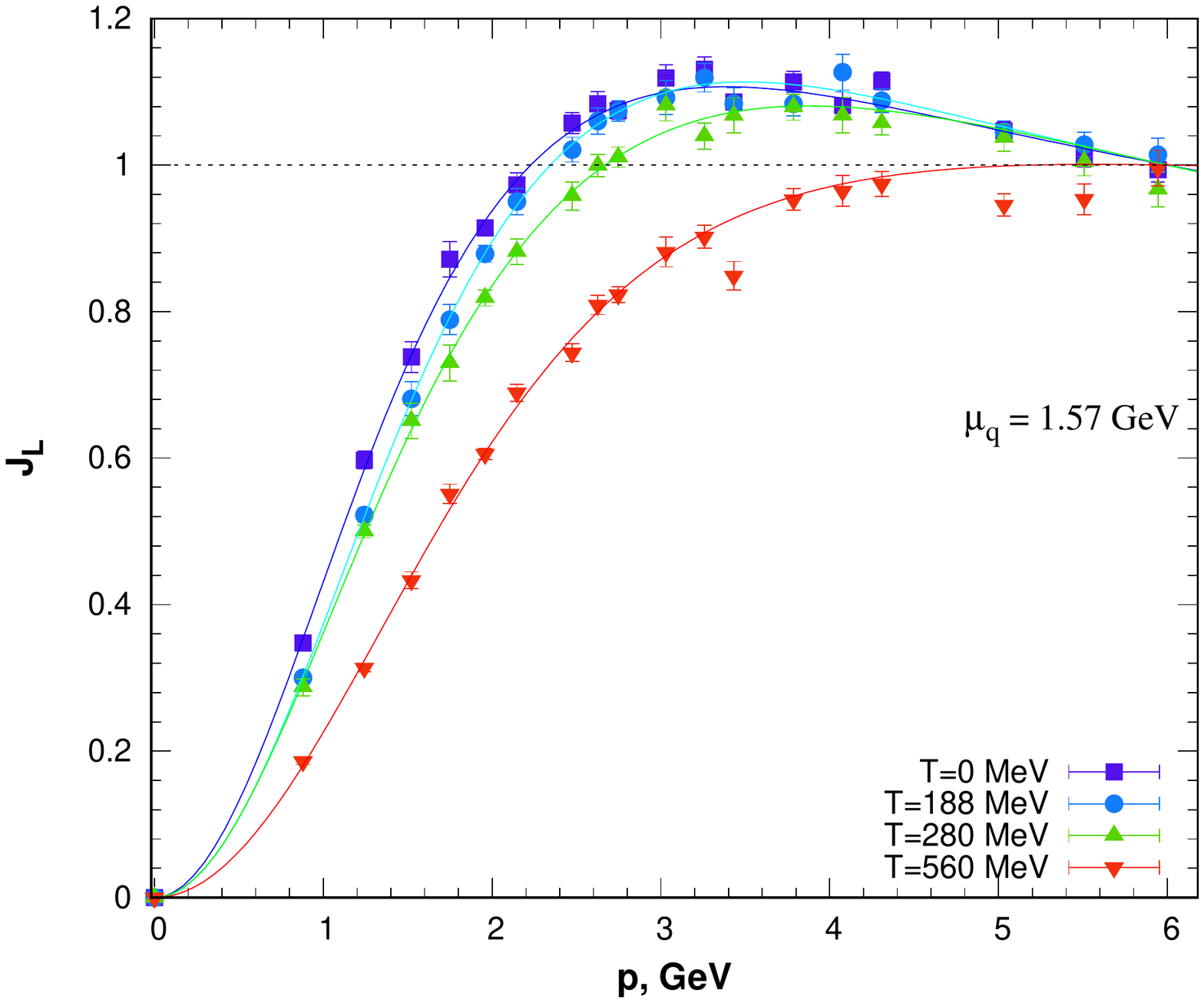}
\includegraphics[scale=0.4]{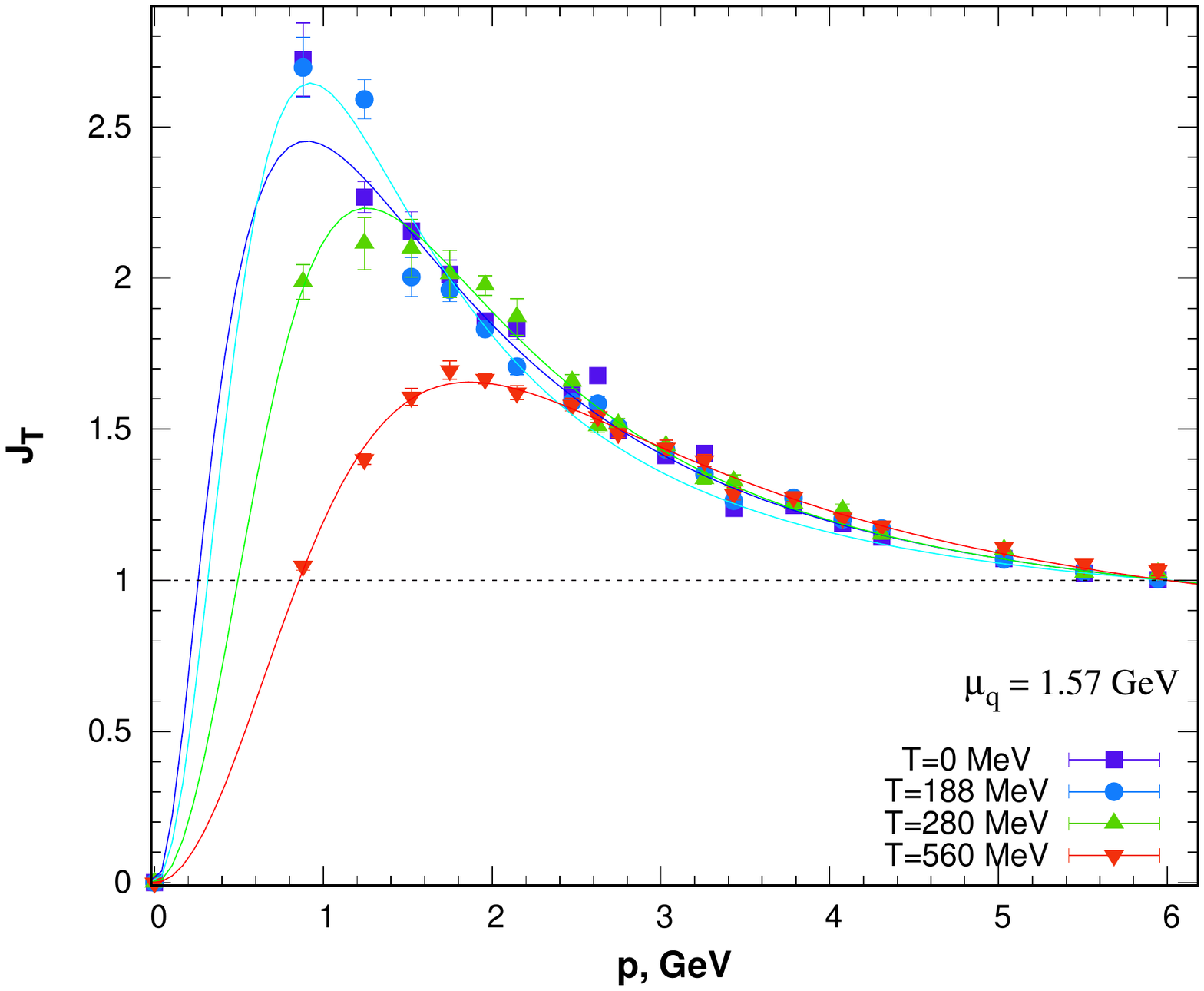}
\vspace*{-35mm}
\caption{Longitudinal (left column) and transverse (right column) gluon dressing functions
at $\mu_q =0.00$ (first row), $\mu_q =0.90$~GeV (second row), and $\mu_q =1.57$~GeV (third row)    for four temperatures.}
\label{fig:dressing_xx_GS}
\end{figure}

The gluon dressing functions 
\beq
J_{L,T}(p)=p^2 D_{L,T}(p) 
\eeq
as well as the
curves derived from the Gribov-Stingl fit (\ref{eq:GS_fit_widely_used}) 
are shown in Fig.\ref{fig:dressing_xx_GS}. 
It should be emphasized that the zero-momentum 
propagator values were taken into account in the fit;
the respective information  is encoded  
in the behavior of the dressing functions at $p\to 0$.

In general, a shape of the dressing functions
is similar to that in pure gluodynamics at finite temperature; 
specific features of their behavior are as follows.

When we increase either $T$ or $\mu_q$ (or both)
the value of the momentum at which the
{\bf longitudinal} dressing function $J_L$ has a maximum increases,
the width of the peak also increases, while the maximal value 
of the dressing function decreases.  
At high $T$ and/or $\mu_q$, $J_L$
tends to that of a free massive particle;
at $T=560$~MeV and $\mu_q=1.57$~GeV it comes close to 
the dressing function of a free particle of mass
$m\approx 2$~GeV.

A completely different situation  is observed
in the {\bf transverse} case. At $T=560$~MeV the dressing function
is practically independent of $\mu_q$; at lower temperatures 
the peak position does not depend on $\mu_q$,
whereas the peak value moderately increases with increasing $\mu_q$,
the width of the peak depends only weakly on $\mu_q$.
Therewith, an increase of the temperature at a given chemical
potential gives {\it (i)} an increase of the peak position, 
{\it (ii)} a decrease of the peak value,
{\it (iii)} a moderate increase of the peak 
width. The transverse gluon dressing functions at all temperatures under study coincide at momenta above $\approx 2$~GeV for $\mu_q=0$ and at momenta above $\approx 2.5$~GeV for high $\mu_q$ values. This is an indication of smallness of the magnetic screening mass at all values of $T$ and $\mu_q$.  

\subsection{Temperature dependence at various momenta and 
quark chemical potentials}

Both propagators change only a little 
when the temperature varies below 280~MeV,
whereas they show a significant decrease as the temperature
increases from $T=280$~MeV to $T=560$~MeV.

\begin{figure*}[hhh]
\vspace*{-20mm}
\hspace*{-8mm}\includegraphics[scale=0.35]{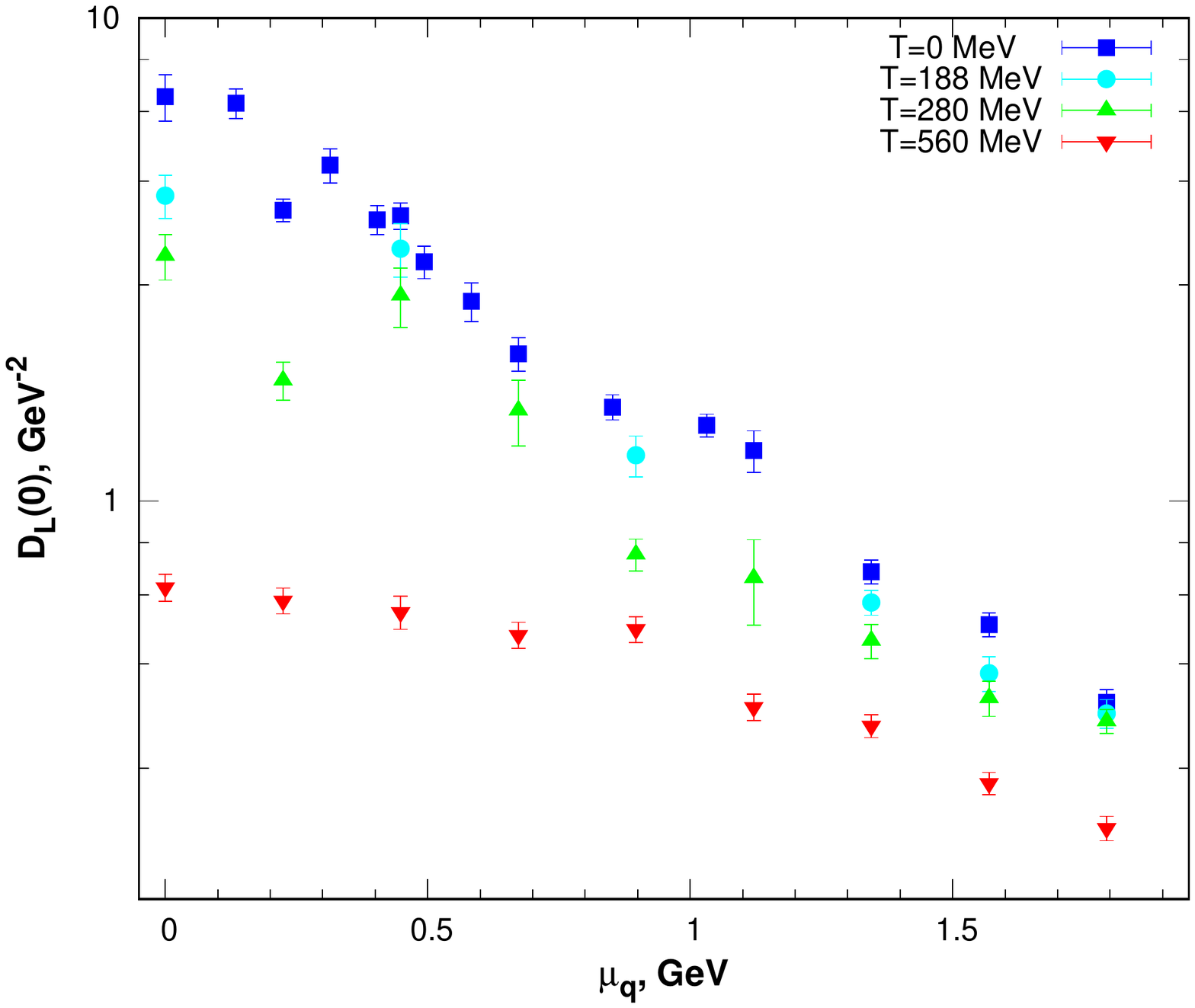}
\includegraphics[scale=0.35]{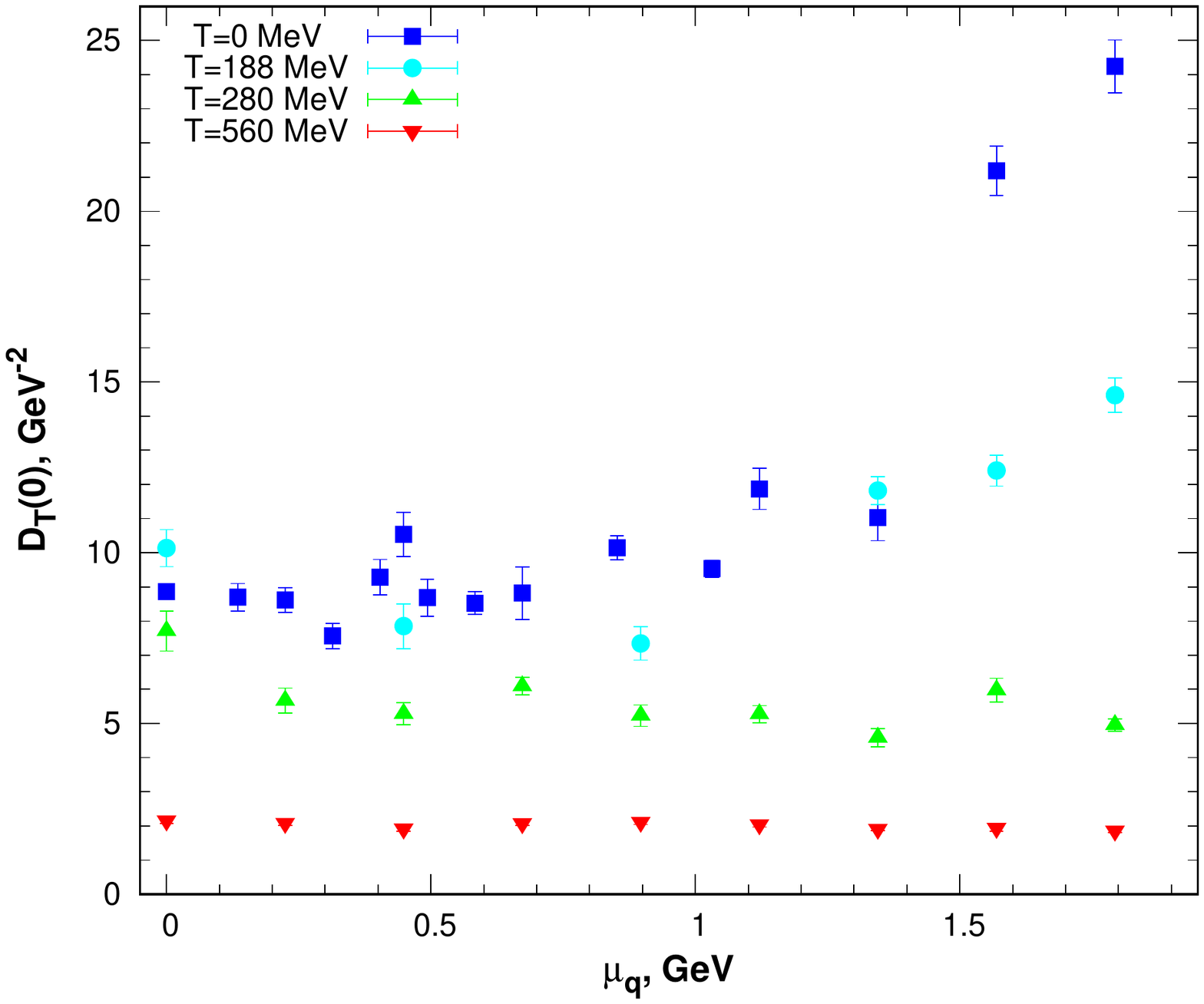}

\vspace*{-40mm}
\hspace*{-8mm}\includegraphics[scale=0.35]{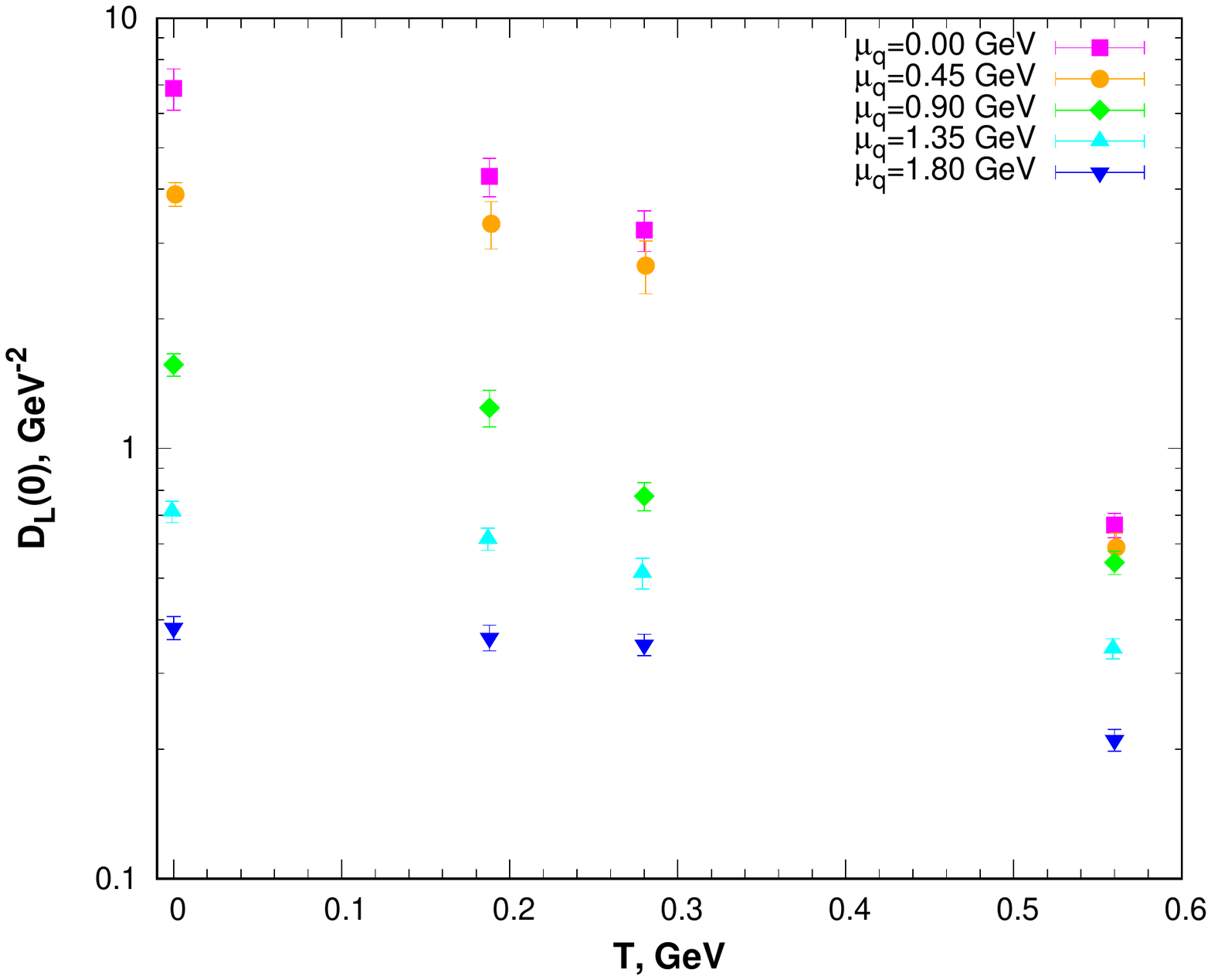}
\includegraphics[scale=0.35]{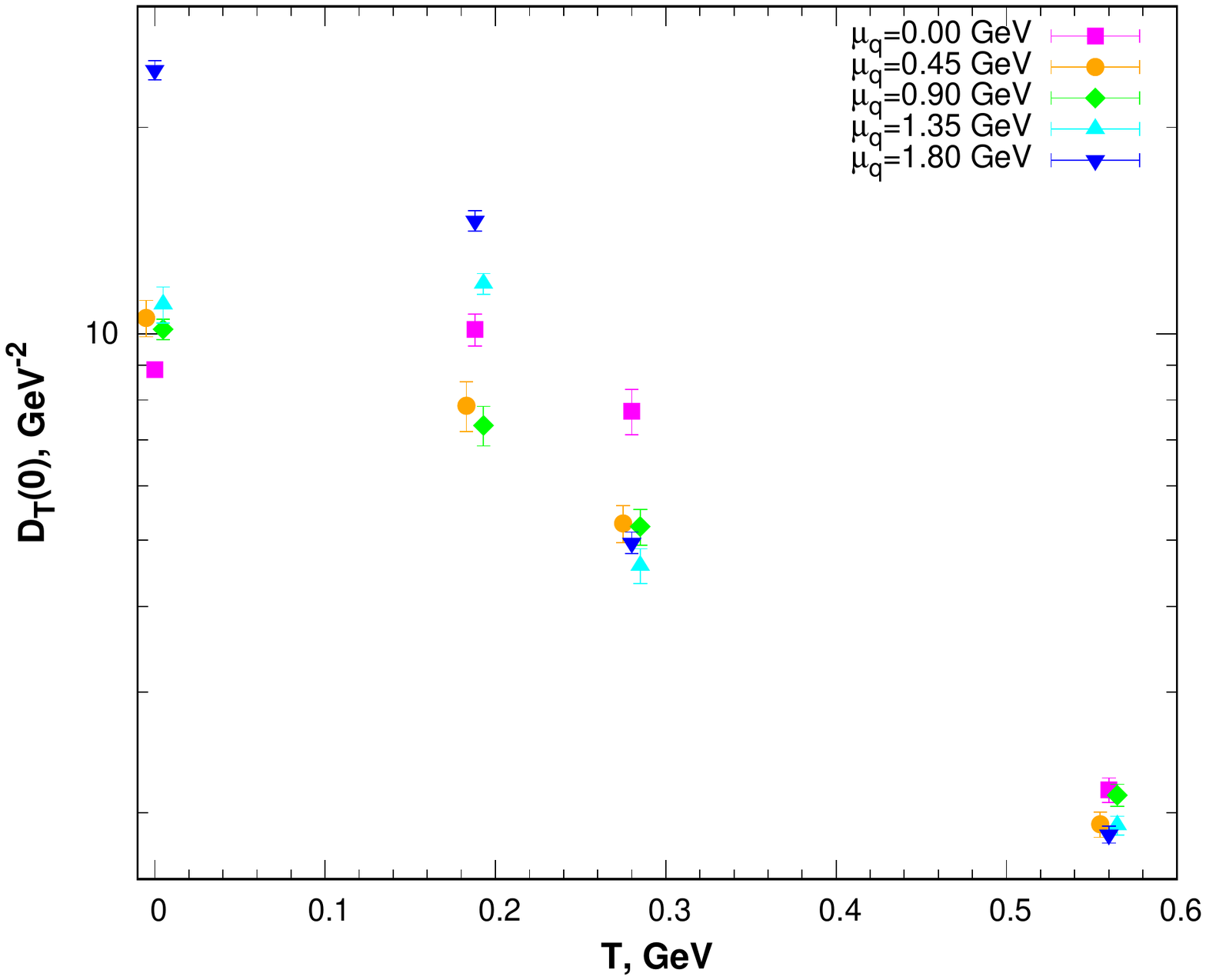}
\vspace*{-30mm}
\caption{Zero-momentum longitudinal (left column) and transverse (right column) gluon propagators
as functions of $\mu_q$ at various temperatures (upper row) and as functions of $T$ at various $\mu_q$ (lower row).}
\label{fig:Dz_vs_Tmu}
\end{figure*}

This being so, temperature dependence of 
{\bf the longitudinal propagator}
is substantial at $p \lesssim 4$~GeV at small $\mu_q$ 
and it is seen up to the normalization point $p=6$~GeV 
at high $\mu_q$. In the infrared, the longitudinal 
propagator decreases significantly with the temperature 
at $\mu_q\lesssim 1.2$~GeV, whereas at greater values
of $\mu_q$ it shows a less pronounced decrease. This can be seen also in Fig.~\ref{fig:Dz_vs_Tmu}~(left, upper row) where dependence of the propagators at zero momentum on temperature and quark chemical potential are depicted. The value of $D_{L,T}(0)$ is of special importance since it is used in one of the definitions of the screening mass, see, e.g. review 
Ref.~\cite{Maas:2011se}.

Temperature dependence of {\bf the transverse propagator}
is seen only at relatively low momenta ($p \lesssim 2\div 2.5$~GeV) as was already said above. Similar to the longitudinal case, it decreases with
temperature. However, in contrast to the longitudinal
propagator, its variation with the temperature 
increases with increasing $\mu_q$ (see also Fig.\ref{fig:Dz_vs_Tmu}~(right)). 


\subsection{The dependence on the quark chemical potential at various momenta and temperatures}

{\bf The longitudinal propagator} gradually decreases with the chemical potential at all momenta ($p \leq 6$~GeV) and temperatures  under consideration. The decreasing is more pronounced at low momenta and temperatures. 

{\bf The transverse propagator} at low momenta and temperatures ($T<200$~MeV, $p \lsim 2\div 2.5$~GeV) and  $\mu_q\gsim 1$~GeV increases with increasing $\mu_q$. 
At $T=280$~MeV its dependence on $\mu_q$ becomes less pronounced and it disappears completely at $T=560$~MeV.
At high momenta the transverse propagator is independent of $\mu_q$ at all temperatures.

\section{Screening masses\\ and interaction potentials}

\subsection{Definition of the screening mass}

In the general case, a widely used definition of the screening mass stems from the on-mass-shell
renormalization of the propagator: the inverse propagator is considered as a regular function 
in some neighborhood of $p^2=0$ and thus represented in the form
\beq\label{eq:OMS_renorm}
D^{-1}_{L,T}(p) = Z ^{-1}(m_{E,M}^2+\tilde \Pi(p^2) + p^2)\;,
\eeq
where the Taylor expansion of the function $\tilde \Pi(p^2)$
starts from the order $\underline{O}((p^2)^2)$ terms.
The propagator of renormalized fields $A_R=Z^{-\,1/2}A$ has the form
\beq
D_{L,T}^{ren}(p) = {1\over m_{E,M}^2 + p^2 + \tilde \Pi(p^2)} 
\eeq
and, if $\tilde \Pi(p^2)$ is small in the infrared, it has a pole
at $p^2\approx -m_{E,M}^2$.
Thus the parameter  $m_{E,M}$ can be associated both 
with the mass of the particle and with the asymptotic behavior
of the propagator at spatial infinity 
\beq\label{eq:glp_large_x_as}
D_{L,T}(0,\vec x)\simeq C_{E,M} e^{-m_{E,M}|\vec x|}\; .
\eeq

In Refs.~\cite{Bornyakov:2010nc,Oliveira:2010xc,Silva:2013maa}
the chromoelectric and chromomagnetic screening masses
were determined using the Yukawa-type fit function
\beq\label{scrmass_mom1}
D_{L,T}^{-1}(p) = Z_{E,M}^{-1}(m_{E,M}^2 + p^2)
\eeq
at zero and finite temperatures.
It was shown \cite{Silva:2013maa} that the Yukawa-type fit function (\ref{scrmass_mom1}) provides a good quality of this fit over rather wide range of momenta 
giving evidence for smallness of $\tilde \Pi(p^2)$ in the infrared. 

The above definition of $m_{E,M}^2$ can be related to the correlation length:
\beq
\label{mass_def1}
m_{E,M}^2=\xi_{E,M}^{-2},
\eeq
where the  correlation length $\xi_{E,M}$ is conventionally defined in terms of the correlation function (propagator in our case)  by the expression \cite{ShangKengMa}
\beq
\label{eq:correlation}
 \xi^2 = \frac{1}{2} {\int_V dx_4 d\vec x  \tilde D(x_4, \vec x) |\vec x|^2
\over  \int_V dx_4 d\vec x   D(x_4, \vec x)} =
- {1\over 2 D(0, \vec 0)}\; \sum_{i=1}^3
\left({d\over dp_i}\right)^2\Big|_{\vec p=0} D(0, \vec p)\; .
\eeq

We consider definition of the screening masses based on the fit of $D_{L,T}^{-1}(p)$ 
at low momenta to a polynomial in $p^2$:
\beq\label{eq:LOW_MOM_fit_fun}
D_{L,T}^{-1}(p) = Z_{E,M}^{-1}\big( m_{E,M}^2 + p^2 + c_4 \cdot (p^2)^2 +... \big)\; .
\eeq
This method was used in Refs.~\cite{Bornyakov:2011jm,Bornyakov:2019jfz}. 
However, we use the function (\ref{eq:LOW_MOM_fit_fun}) rather than function (\ref{scrmass_mom1})  
because we have no enough data points in the infrared region where the propagator can be described by the function
(\ref{scrmass_mom1}). Thus, to obtain a reasonable fit results we had to use terms up to $(p^2)^2$ for $D_L(p)$ and
terms up to $(p^2)^3$ for $D_T(p)$. Still, we hope that 
making use of the fit function (\ref{eq:LOW_MOM_fit_fun})
provides reasonably good estimates of the parameters in eq.~(\ref{scrmass_mom1}).

When the screening masses are large, it is natural to assume 
that the one-gluon exchange dominates. In the
approximation of one-gluon exchange, the potential 
of chromoelectrostatic  interaction between static
external color sources is given by the Fourier
transform of the longitudinal propagator 
(the  chromomagnetostatic potential between currents 
- by the Fourier transform of the transverse
propagator). 
Both chromoelectric and chromomagnetic potentials are
short-range; that is, they can be roughly described
by a potential well of certain width and depth.
The parameter characterizing the width is provided
by $\ds \xi_{E,M}={1\over m_{E,M}}$, whereas 
the parameter $V_{E,M}$ characterizing the depth
can be defined as follows:
\beq
\int d\vec x{\cal V}_{E,M}(\vec x)=D_{L,T}(p_4=0,\vec p=0)=
 V_{E,M} \xi_{E,M}^3 
\eeq
\beq
\implies \qquad 
V_{E,M}= D_{L,T}(0) m_{E,M}^3\; ,
\label{VEM1}
\eeq
where ${\cal V}_{E,M}(\vec x)$ is the chromoelectric 
(chromomagnetic) potential
describing interaction between static color charges (currents).

Before proceeding further, we recollect again another definition of the screening mass 
\cite{Maas:2011se}:
\beq
\label{maas}
{\cal M}_E^2={1\over D_L(0)}\;, \qquad  \; {\cal M}_M^2={1\over D_T(0)}.
\eeq
Clearly, it depends on renormalization and is rather sensitive to the finite volume effects.
Loosely speaking, eq.~(\ref{maas})  characterizes ``the total amount'' of the interaction since
\beq
{1\over {\cal M}_{E,M}^2} = \int dx_4 d\vec x D_{L,T}(x_4, \vec x) = \int d\vec x\; {\cal V}_{E,M}(\vec x) \; ,
\eeq
where $D_{L,T}(x_4,\vec x)$ are the propagators in the coordinate representation.

It follows from (\ref{VEM1}) and (\ref{eq:LOW_MOM_fit_fun}) that
\beq\label{eq:V_from_Z}
V_{E,M}=Z_{E,M} m_{E,M}\; ,
\eeq
where $Z_{E,M}$ is the respective renormalization factor. The numerical results for $V_{E,M}$
will be presented below.

\subsection{Dependence of screening masses and interaction potentials on $\mu_q$ and $T$}
\begin{figure*}[hhh]
\vspace*{-17mm}
\hspace*{-11mm}\includegraphics[width=7.5cm]{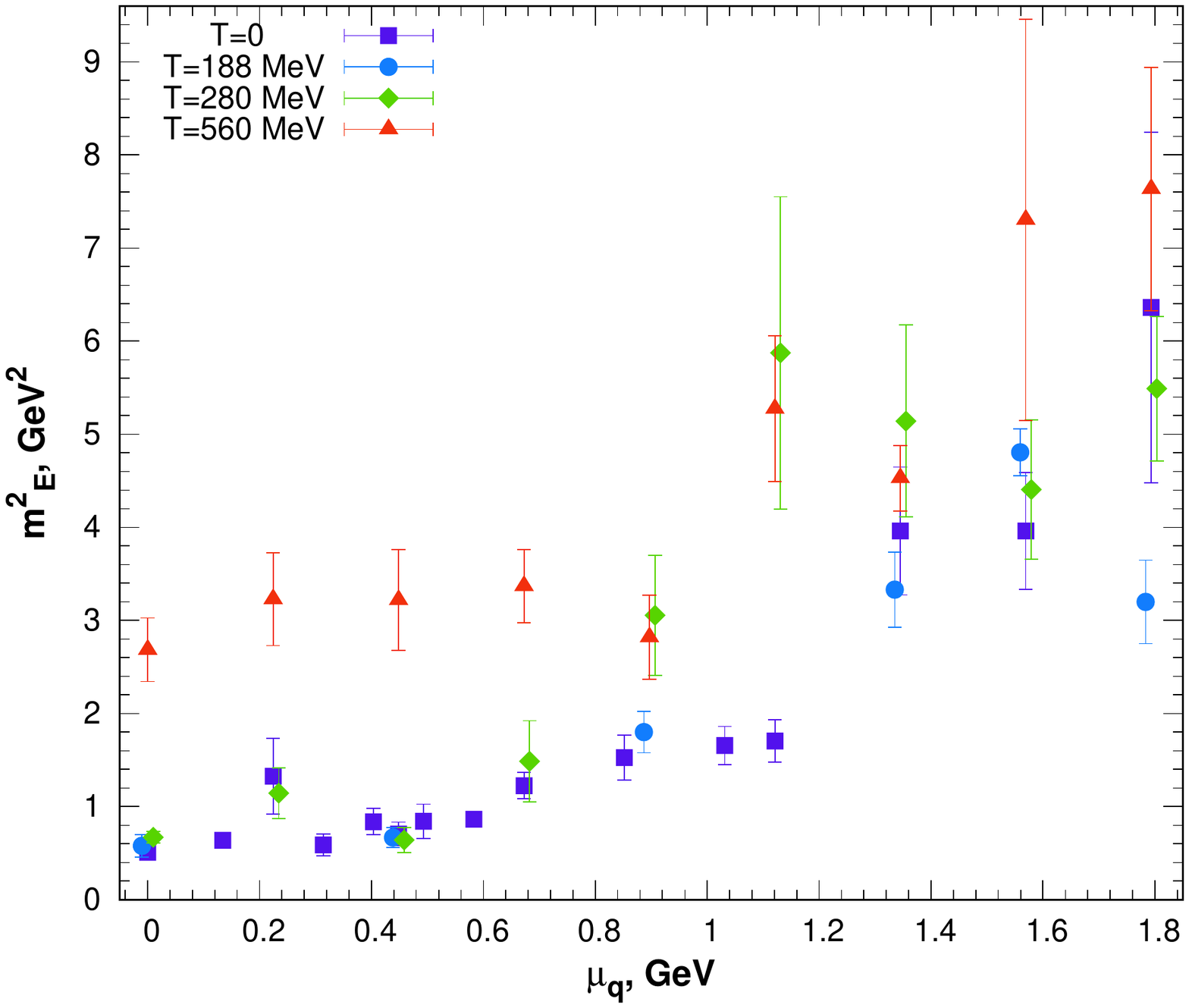} \includegraphics[width=7.5cm]{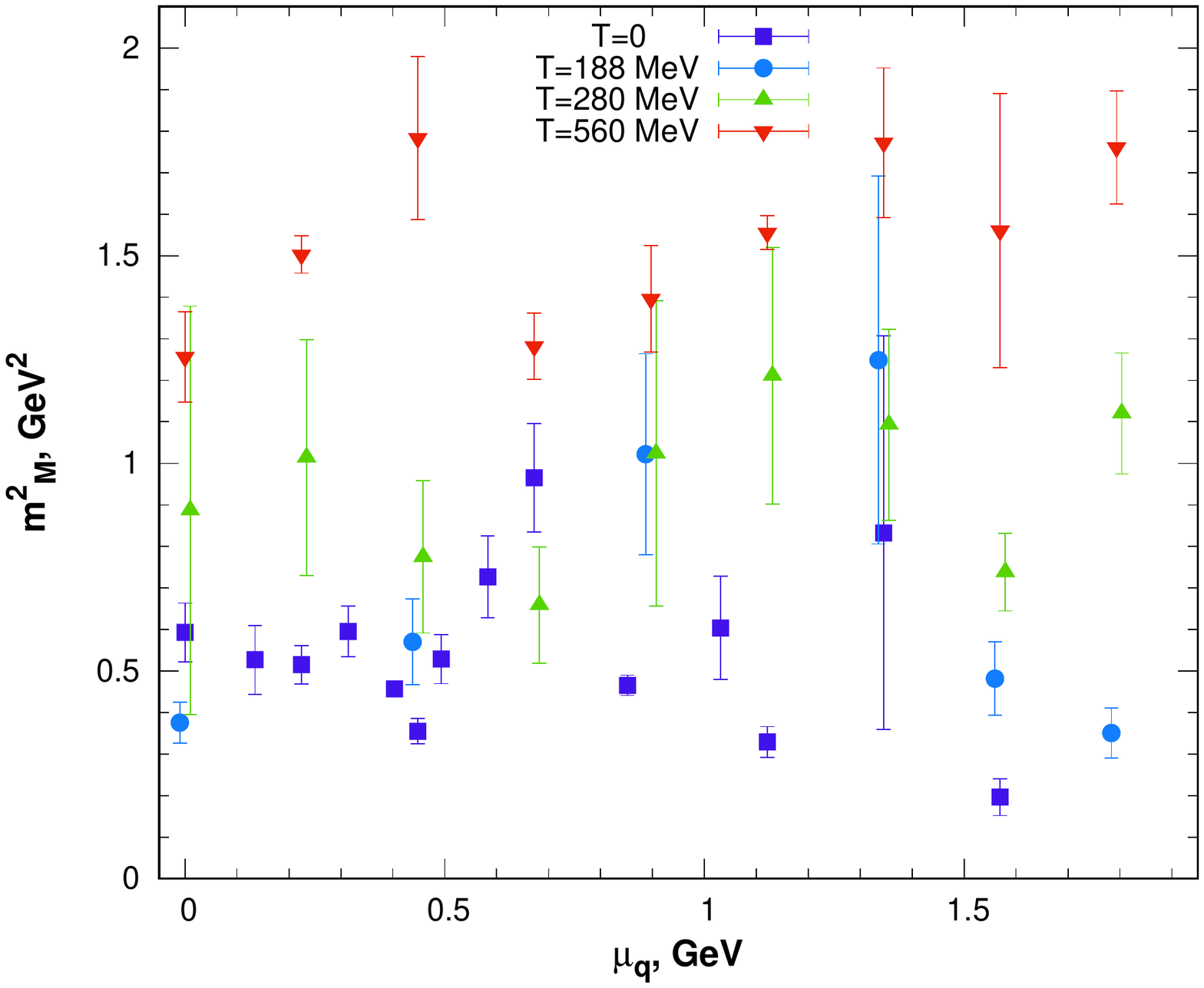}
\vspace*{-28mm}
\caption{Chromoelectric (left) and chromomagnetic (right)
screening masses 
as functions of $\mu_q$ at various temperatures.
}
\label{fig:mass_vs_mu}
\end{figure*}

In Fig.\ref{fig:mass_vs_mu} we show the electric (left panel) and magnetic (right panel) masses as functions of the quark chemical potential at various temperatures.


Our value for $m_E/\sqrt{\sigma}$ at $\mu_q=0$ and $T=0$ is 1.50(4) \cite{Bornyakov:2019jfz}. This value should be compared with the value 1.47(2)
obtained in SU(3) gluodynamics at zero temperature \cite{Oliveira:2010xc}   by fitting the
inverse propagator to the form  (\ref{scrmass_mom1})  at small momenta\footnote{We obtained this value taking
mass value 647(7) MeV, obtained in \cite{Oliveira:2010xc} and dividing it by $\sqrt{\sigma} = 440$MeV used
in \cite{Oliveira:2010xc} to set the scale.}. We also quote a value  1.48(5) obtained for a mass dominating the small momentum behavior of a gluon propagator in  $SU(2)$ lattice gluodynamics in \cite{Langfeld:2001cz}.

At temperatures $T\leq 300$~MeV,  $m_E$ does not significantly change at small $\mu_q$ corresponding to the hadron phase. Above $\mu_q \approx 200$~MeV it starts to increase and continues to increase up to 1.8 GeV.
This behavior is analogous to that of the electric 
screening mass in QCD at $T>T_c$ as was demonstrated 
by lattice simulations
\cite{Fischer:2010fx,Maas:2011ez,Silva:2013maa,Bornyakov:2010nc}.
No such increasing was reported in the earlier 
studies\cite{Boz:2018crd} of QC$_2$D.

\begin{table}[h]
\begin{center}
\vspace*{0.5cm}
\begin{tabular}{|c|c|c|} \hline
                      &  $\mu_q<850$~MeV     &  $\mu_q > 850$~MeV  \\ \hline\hline
$T< 300$~MeV      &   $m_E\simeq 0.7$~GeV             &  $m_E\simeq 1.8\div 2.4$~GeV  \\
                       &   $m_M\simeq 0.7$~GeV             &  $m_M\simeq 0.4\div 1.1$~GeV  \\ \hline
$T=560$~MeV       &    $m_E\simeq 1.6\div 1.8$~GeV    &  $m_E\simeq 2.4\div 3.0$~GeV  \\
                       &  $m_M\simeq 1.2\div 1.3$~GeV     &  $m_M\simeq 1.3\div 1.5$~GeV  \\
\hline\hline
\end{tabular}
\end{center}
\caption{Dependence of the screening masses on the quark chemical potential and temperature.
It should be emphasized that, at $T=0$ and $\mu_q > 850$~MeV, the magnetic mass decreases from $m_M\simeq 700$~MeV to $m_M\simeq 400$~MeV.}
\label{tab:mE_and_mM_vs_muq_and_T}
\end{table}

Information on the behavior of the screening masses 
is summarized in the table.
Chromoelectric forces feature the longest range 
at $T=0$ and $\mu_q=0$; their screening increases 
with an increase of both $T$ and $\mu_q$.

Chromomagnetic forces have the same radius 
as chromoelectric at $T=0$ and $\mu_q=0$;
however, their screening increases with $T$, 
whereas $\mu_q$-dependence of their screening
depends on the temperature: $m_M$ decreases with $\mu_q$ 
at low temperatures, remains constant at $T=280$~MeV 
and tends to increase at $T=560$~MeV.

It should be emphasized that an incerase of $m_M$ is clearly seen provided that high temereatures are taken into consideration. No such increase was observed in Ref.~\cite{Boz:2018crd} where the temperature range $T<1.6 T_c$ was investigated; in this temperature range our data tend to confirm the previous results: a very limited growth of $m_M$ is observed at $T< 280$~MeV. 

It should also be noticed that the $Z$ factors obtained here
differ from the ratio
\beq\label{eq:eta}
 \eta_{E,M}(\mu_q) = { m_{E,M}^2(\mu_q)\over {\cal M}_{E,M}^2(\mu_q) }  = 
m_{E,M}^2(\mu_q) D_{L,T}(0)
\eeq
considered in our previous study \cite{Bornyakov:2019jfz}. The reason is as follows: we obtain $Z$ from the
fitting procedure, which is in fact equivalent to making use of the formula (\ref{eq:eta}),
however, with $D_{L,T}(0)$ evaluated as zero-momentum value of the fit function (\ref{eq:LOW_MOM_fit_fun}) rather than the value extracted from data immediately.

In our previous study \cite{Bornyakov:2019jfz} devoted to zero-temperature case, $Z_E$ was considered as independent of $\mu_q$ 
apart from variation at small $\mu_q$ values.
However, in the present study we found that 
both $Z_{E}$ and $Z_{M}$ depend either on $\mu_q$ or on $T$.

\begin{figure}[tbh]
\vspace*{-17mm}
\hspace*{-10mm}\includegraphics[width=7.2cm]{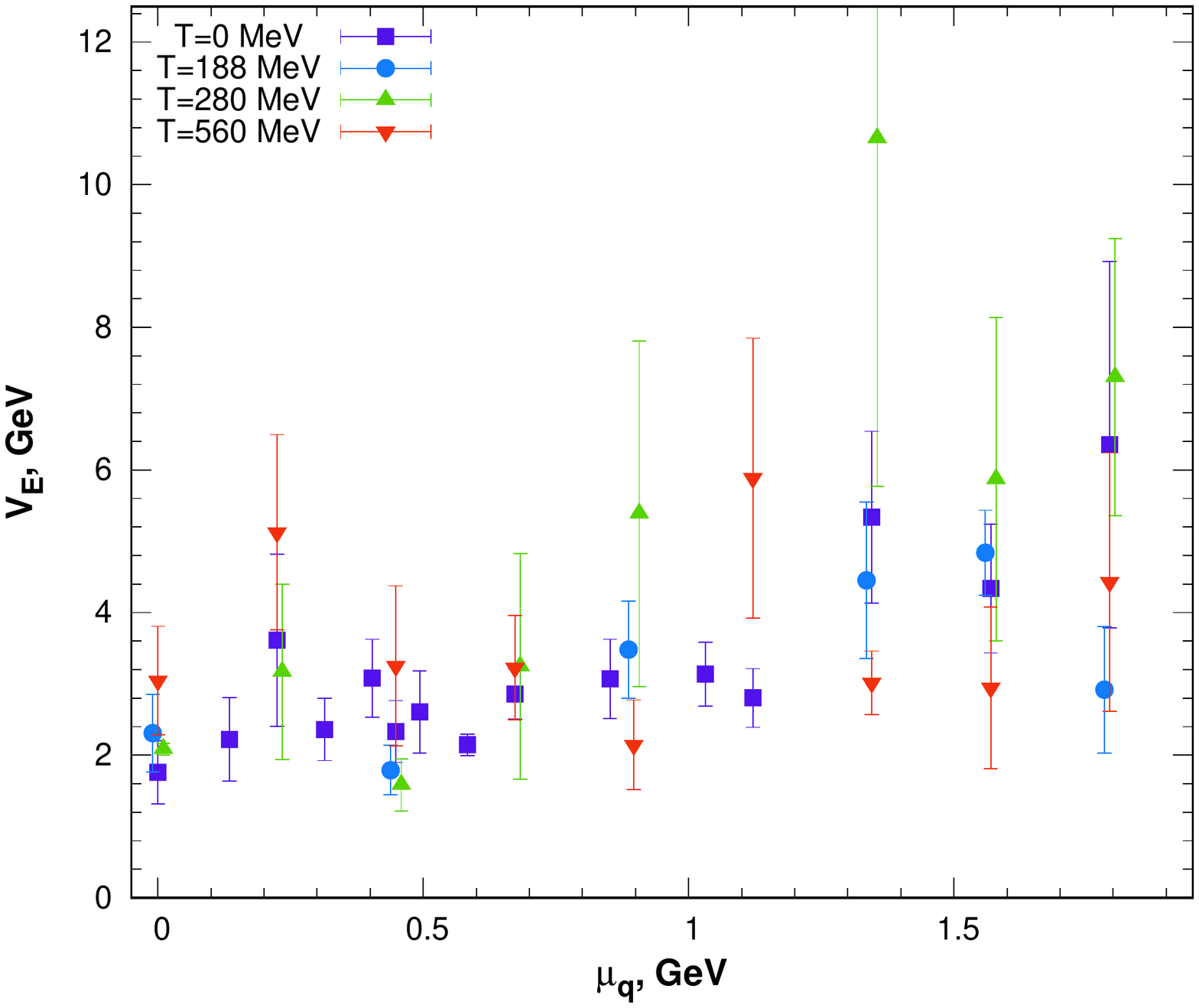} \includegraphics[width=7.2cm]{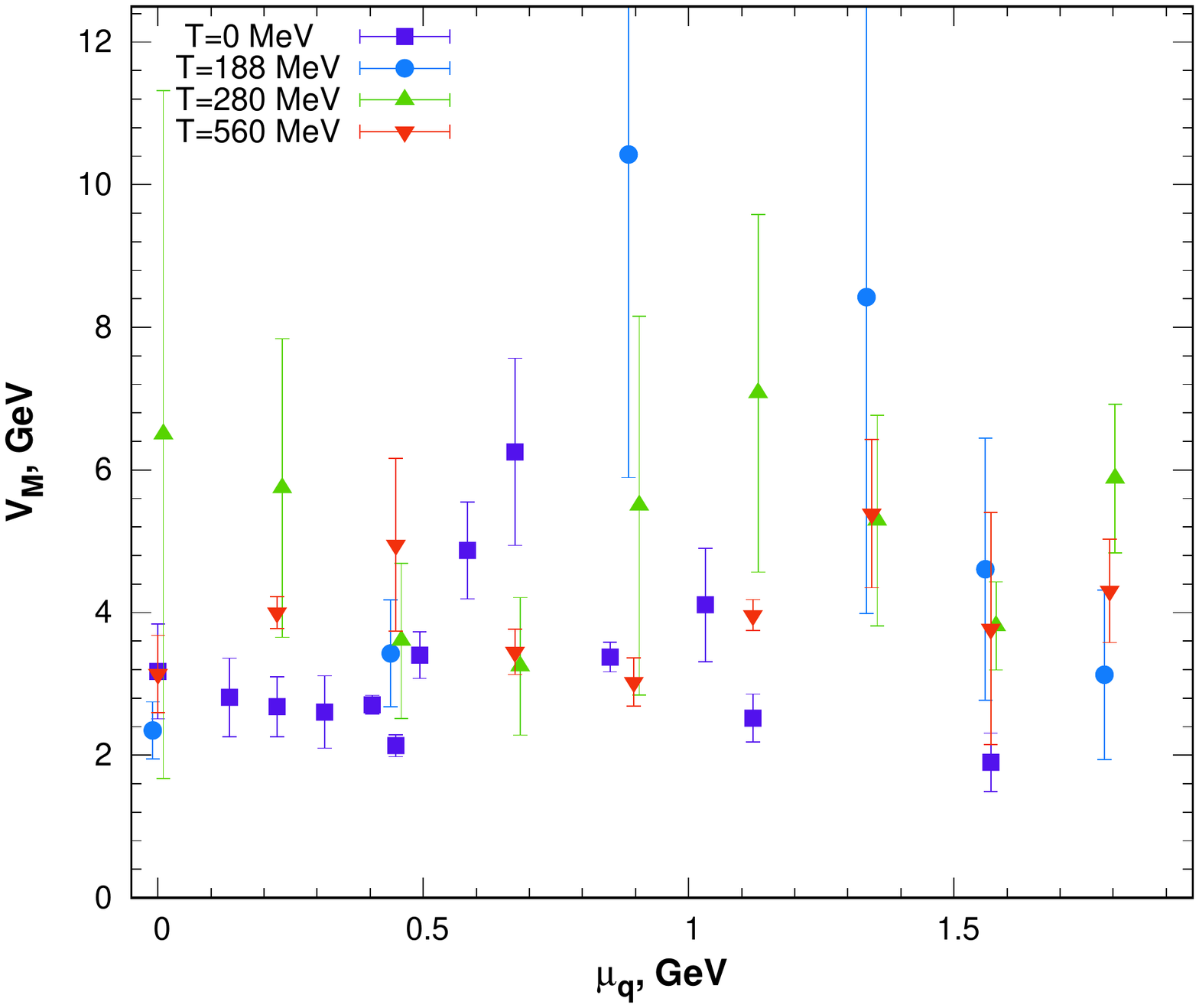}
\vspace*{-24mm}
\caption{The parameter $V_{E,M}$ characterizing interaction
between color static sources as functions of the quark chemical potential at
various temperatures. Huge errors make only qualitative estimates possible.
}
\label{fig:V_vs_mu_n_T}
\end{figure}

The depth of the potential well can be roughly estimated 
from the formula (\ref{eq:V_from_Z}), the results are shown in Fig.\ref{fig:V_vs_mu_n_T}. It varies over the range $2\lesssim V_{E,M}\lesssim 6$~GeV.
Huge errors do not allow to perform quantitative comparison
of $V_E$ with $V_M$, as seen by eye, $V_M$ tends 
to be a little greater.

\section{Conclusions}
\begin{itemize}
\item We investigated the dependence of $D_T$ and $D_L$ 
on the temperature and isospin chemical potential.
Both propagators decrease with increasing temperature;
however, they behave differently as the functions of $\mu_q$. $D_L$ decreases with increasing $\mu_q$ at all temperatures under study, whereas $D_T$ increases with increasing $\mu_q$ at $T<200$~MeV and is independent of $\mu_q$ at higher temperatures.

\item In the model under study, the radius of action of the chromomagnetic forces is greater than that of chromoelectric forces at all temperatures and chemical potentials
excepting a neighborhood of the point $\mu_q=0, T=0$.
Within the range of action, the strength of chromomagnetic forces is also approximately the same or a little greater than that of chromoelectric ones (at $\mu_q>0, T>0$). For this reason, the 
strong interacting matter described by $QC_2D$  at nonzero
temperatures and quark chemical potentials can be named a chromomagnetic medium.

\end{itemize}

\section*{Acknowledgments}
The work was completed due to support of the Russian Foundation for Basic Research via grant 18-02-40130 mega.
The authors are thankful to Victor Braguta, Andrey Kotov and Alexander Nikolaev for providing gauge field configurations and useful discussions. 
The research is carried out using the Central Linux Cluster of the NRC ``Kurchatov Institute'' - IHEP, the equipment
of the shared research facilities of HPC computing resources at Lomonosov Moscow State University, the Linux Cluster
of the NRC ``Kurchatov Institute'' - ITEP (Moscow). In addition, we used computer resources of the federal collective
usage center Complex for Simulation and Data Processing for Mega-science Facilities at NRC Kurchatov Institute,
http://ckp.nrcki.ru/.

\end{document}